\documentclass[12pt,preprint]{aastex}

\begin{document}

\title {Optical observations of type IIP supernova 2004dj:
Evidence for the asymmetry of $^{56}$Ni ejecta}

\bigskip

\author{N.N.~Chugai\altaffilmark{1}, S.N.~Fabrika\altaffilmark{2}, 
 O.N.~Sholukhova\altaffilmark{2}, V.P.~Goranskij\altaffilmark{2},
P.K.~Abolmasov\altaffilmark{3}, and V.V.~Vlasyuk\altaffilmark{2}
}
\affil{$^{1}$Institute of astronomy RAS, Moscow}
\affil{$^{2}$Special astrophysical observatory, Zelenchuk}
\affil{$^{3}$Sternberg astronomical institute, Moscow}

\begin{abstract}

The photometric and spectroscopic observations 
of nearby type IIP supernova 2004dj are presented. 
The $^{56}$Ni mass estimated from the light curve
is $\approx0.02~M_{\odot}$. This estimate is found to be 
consistent with the H$\alpha$ luminosity.
SN2004dj reveals a strong asymmetry of the H$\alpha$ emission line 
at the nebular epoch with the shift of the maximum of
 $-1600$ km s$^{-1}$.
A similar asymmetric component is detected in H$\beta$, 
[O I] and [Ca\,II] lines. The line asymmetry is 
interpreted as a result of the asymmetry of $^{56}$Ni ejecta.
The H$\alpha$ profile and its evolution are reproduced in 
the model of the asymmetric bipolar $^{56}$Ni and 
spherical hydrogen distributions.
The mass of the front $^{56}$Ni jet is 
 comparable to the central component and 
 twice as larger compared to the rear $^{56}$Ni jet.
We note that the asymmetric bipolar structure 
of of $^{56}$Ni ejecta is revealed also by SN1999em, 
another type IIP supernova.

\end{abstract}

\newpage

\section{Introduction}

Supernova SN2004dj is discovered on 2004 July 31 in Sc galaxy 
NGC2403 (Nakano et al. 2004). 
According to spectra and light curve (Korcakova 2005) this 
is a usual type IIP supernova (i.e., with plateau in the light curve)
detected in 30-45 days after the explosion.
With the distance of 3.13 Mpc (Freedman et al. 2001) this 
is closest SNIIP after SN1987A. Given the peculiar nature of 
the latter SN2004dj turns out to be closest ever normal SNIIP 
observed.

Type IIP supernovae are related with the explosion 
of a massive ($10-25~M_{\odot}$) red 
supergiant following the collapse of iron core.
Three versions of the explosion mechanism have been proposed:
(1) neutrino mechanism (Colgate \& White 1966; Buras et al. 2003);
(2) magneto-rotational mechanism (Ardelyan et al. 2005); 
(3) rotational fragmentation of a new-born neutron star with the 
subsequent explosion of low-mass ($\sim0.1~M_{\odot}$) fragment 
(Blinnikov et al. 1990). Whatever happens the subsequent shock wave
ejects the envelope with the sufficient internal energy to provide
the SNIIP luminosity at the plateau stage (Grasberg et al. 1971).
The shock wave propagation in the silicon mantle results in the 
synthesis of radioactive $^{56}$Ni; its decay via 
$^{56}$Ni--$^{56}$Co--$^{56}$Fe maintains the SNIIP luminosity 
at the tail of the light curve (Weaver \& Woosley 1980). 
Neither of proposed explosion mechanism is able as yet to reproduce 
empirical values of the explosion energy and ejected $^{56}$Ni mass
of SNIIP.

The proximity of SN2005dj provides us an opportunity to study 
in greater details 
the observational display of an explosion of normal SNIIP.
Unfortunately, supernova was discovered long after the 
outburst so the explosion energy cannot be recovered 
from a hydro modeling.
Yet the light curve and H$\alpha$
luminosity at the nebular epoch can provide us an estimate 
of the $^{56}$Ni mass, 
another vital property of the explosion mechanism.

Of particular interest would be the emission line profiles 
at the nebular epoch, which 
could provide us valuable information about the possible
 explosion asymmetry in SNIIP. The asymmetry issue 
 emerged after the detection of the emission lines 
 redshift in SN1987A 
 (Phillips \& Williams 1991) followed by the interpretation of this 
 redshift as an effect of $^{56}$Ni ejecta asymmetry (Chugai 1991). 
A similar phenomenon was recovered recently in 
 the type IIP supernova SN1999em (Elmhamdi et al. 2003a). 
Herant et al. (1992) conjecture that the $^{56}$Ni asymmetry 
as well as 
high pulsar velocities may be related to the violation of a 
point symmetry in core-collapse supernovae 
owing to the large scale neutrino convection.

Below we present results of spectral and photometric 
observations of SN2004dj in Special astrophysical observatory (SAO) 
at Zelenchuk. Apart from a qualitative analysis of the 
observational data we present estimates of  $^{56}$Ni mass in
SN2004dj from the light curve and the H$\alpha$ luminosity.
The observed spectra reveal unusually strong asymmetry of  
the H$\alpha$ emission line. Given the importance of this 
phenomenon we concentrate on the study of the line asymmetry 
and its relationship with the asymmetry of $^{56}$Ni ejecta. 

Henceforth we use the distance $D=3.13$ Mpc (distance module $\mu=27.48$)
according to cepheids (Freedman et al. 2001) and the redshift of
129 km s$^{-1}$ according to LEDA database.

\section{Observations}

Broad band $BVRc$ photometry of SN2004dj was carried out at 
1-m Zeiss telescope of SAO using CCD Electronika K-585 and 
EEV 42-40. The photometry data were also obtained at 
60-cm telescope of Crimean station of SAI with CCD VersArray.
These three detectors are shortened as K, E, and VA respectively.
A local photometric standard around supernova is based on the 
$UBVR$ standard in the vicinity of NGC 2403 (Zickgraf et al. 1990).
The reduction of CCD images is made with the WinFITS package
(V.P.~Goranskij) using a method of corrected aperture measurements 
that permits us to deal with 
the inhomogeneous background in the vicinity 
of supernova. The accuracy of the photometric data is 0.005--0.01 mag.
The data are presented in Table 1. The columns in order are 
Julian dates, magnitudes and the detector type. The first line of 
the Table 1 contains the photometry data on 2001 January 19, long before 
the explosion. We believe that these magnitudes refer
 to the star cluster S96 on which SN2004dj is superimposed.

The log of spectral observations which includes observation moments,
spectrograph type, spectral range and resolution is 
presented in Table 2.
Eight spectra were taken between 2004 October 18 and 2005 June 9.
The first spectrum is obtained at the transition between the 
plateau and tail, whereas the rest ones were taken at the light 
curve tail. Three types of spectrograph were used in observations:
(i) Scorpio, the universal focal reducer (Afanasiev \& Moiseev 2005)
of 6-m large azimuthal telescope (LAT), multi-pupil fiber spectrograph
(MPFS) of LAT, and UAGS spectrograph of 1-m telescope of SAO.
CCD images of spectra taken at Scorpio and UAGS are reduced 
using MIDAS package with the usual 
procedures of the extraction of cosmic rays, field flattening, and spectra 
extraction. The reduction of spectra taken 
at MPFS is performed using package written by V.L.~Afanasiev, A.V.~Moiseev 
and P.K.~Abolmasov. In two cases (spectra No. 2 and 6 in Table 2) 
the atmospheric dispersion strongly modifies the spectral energy 
distribution. These spectra will not be used for the H$\alpha$ 
flux measurements, although they may well be used for the line
profile analysis.

\section{Photometry and $^{56}$Ni mass}

The $V$ light curve according to our data and other 
available data$^{1}$ 
\footnotetext[1]{http://www.astrosurf.com/snweb2/2004dj/Meas.htm} 
is plotted in Fig. 1. The time is counted from the adopted 
explosion moment JD2453170 (i.e., 2004 June 13). This choice 
is based  upon the assumption that light curves of SN2004dj and 
SN1999gi, standard SNIIP (Leonard et al. 2002), are 
similar. This assumption is supported by the fact that 
the same fitting curve describes both sets of data 
reasonably well (Fig. 1). 
The approximation suggests the exponential behavior 
of the flux in $V$ band at the tail 
 $F(V)\propto\exp(-t/111.26 \,{\rm d})$ in accordance with the 
radioactive decay of $^{56}$Co--$^{56}$Fe.

The two last measurements (day 266 and 267) show excess 
$\approx 0.17$ mag over the exponential behavior (Fig. 1). 
Most of this excess is likely related to the contribution 
of the star cluster S96. 
Given the pre-explosion magnitude (Table 1) we find that 
the star cluster on day 266.5 must contribute 0.14 mag assuming 
exponential behavior of the intrinsic $V$ magnitude of SN2004dj.
The light of S96, indeed, accounts for, practically, all the excess. 

Galactic reddening towards NGC2403
 is $E(B-V)=0.062$ (Schlegel et al. 1998). This value is 
 adopted as a total reddening of supernova in our and host galaxy.
With this reddening the intrinsic $B-V$ color of 
SN2004dj according to our and other available data 
(see previous footnote) is displayed 
in Fig. 2. We plotted also the SN1987A
color (Catchpole et al. 1987, 1988) and 
the lower envelope of intrinsic colors of SNIIP 
(Elmhamdi et al. 2003b). The $B-V$ color of SN2004dj coincides 
wih the lower envelope of SNIIP colors,  which 
indicates that the adopted reddening is close to the 
actual value. On the other hand Wang et al. (2005) upon the 
basis of the population synthesis of the star cluster 
S96 derive the reddening $E(B-V)\sim0.35$. The contradiction 
may be resolved if the light form S96 is 
absorbed in the cool shell around this cluster, while the
supernova resides outside the shell, closer to the observer.
 
The $V-R$ color of SN2004dj (Fig. 2, inset) around day 150 is 
by 0.3 mag lower than that of SN1987A. However, the color difference 
gets smaller with time and disappears around day 200. 
This indicates that the spectral energy distributions (SED) for both 
supernova in the long wavelength range ($\lambda>5000$ \AA) 
at $t\geq200$ d are similar. This fact is of importance for the 
photometric estimate of $^{56}$Ni mass in SN2004dj.

The amount of $^{56}$Ni in the envelope can be found 
from a comparison of $M_V$ absolute magnitudes of SN2004dj 
and SN1987A at the early nebular epoch ($t<250$ d) when 
the envelope is optically thick for gamma-rays, so one 
may adopt that all the radioactive luminosity is instantly 
re-emitted in the optical. The method assumes that the 
distance and reddening are well known and the bolometric 
correction is similar for both supernovae. 
The latter seems to disagree with the apparent 
difference of their $B-V$ colors. However, one should keep in mind 
that at this epoch the bulk of the energy is emitted in the 
long wavelength range ($\lambda>5000$ \AA), in which 
SED are similar around and after day 200 as indicated by the 
their $V-R$ colors (Fig. 2). Therefore, using SN1987A as a template 
we hopefully can recover the amount of $^{56}$N in SN2004dj 
from $V$ magnitude around day 200. 

Using the exponential approximation of the $V$ light curve of SN2004dj 
(Fig. 1) we get $V=15.32$ on day 200. With the reddening $E(B-V)=0.062$ 
and standard extinction law (Seaton 1979) the absorption is $A_V=0.19$.
With the distance module $\mu=27.48$ for NGC2403 one gets 
 $M_V=-14.01$ on day 200. On the other hand, for SN1987A at this 
 epoch $V=5.09$ (Catchpole et al. 1988). Adopting the distance 
 50 kpc and absorption $A_V=0.6$ one gets 
 $M_V=-14.01$ on day 200. Thus on day 200 SN1987A is brighter 
 in $V$ band compared to SN2004dj by 1.53 mag. 
With the $^{56}$Ni mass $0.075~M_{\odot}$ in the SN1987A envelope
(Suntzeff \& Bouchet 1991) we conclude that the amount of 
$^{56}$Ni in SN2004dj is $M(^{56}{\rm Ni})\approx 0.02~M_{\odot}$.
 
The uncertainty of the $^{56}$Ni mass related to the error in 
distance (Freedman et al. 2001) is of 20\%.
The uncertainty related with the extinction unlikely 
exceeds 10\% taking into account low reddening indicated by 
the extremely "blue" color $B-V$ of SN2004dj. Some contribution 
to the error of $^{56}$Ni mass may be related to the possible difference 
of bolometric corrections of SN2004dj and SN1987A. To summarize, 
the total error of the $^{56}$Ni mass estimate is $\sim30$\%.

Independently, the $^{56}$Ni mass may be estimated using 
the empirical relation between amount of $^{56}$Ni and 
parameter $S=dV/dt$ that characterizes the steepness of the 
$V$ light curve at the transition between plateau and tail 
(Elmhamdi et al. 2003b). In the case of SN2004dj we find 
$S=0.17\pm0.04$ mag. day$^{-1}$. Using the empirical 
relation from (Elmhamdi et al. 2003b) one gets
$M(^{56}{\rm Ni})=0.013\pm 0.004~M_{\odot}$ that is consistent 
within errors with the estimate found above. 

\section{Spectrum and its evolution}

\subsection{Qualitative analysis}

The observed spectra of SN2004dj are presented in Fig. 3. 
The last two spectra (Table 2) are merged into the single 
spectrum since at this epoch the spectral evolution is slow. 
The appearance and evolution of SN2004dj spectrum are similar 
to those of ordinary SNIIP at the nebular epoch. 
The late time spectrum is dominated by emission lines of H$\alpha$,
[O\,I] 6300, 6364 \AA\  and  [Ca\,II] 7291, 7324 \AA\  
on the background of the relatively weak quasicontinuum. 
According to general wisdom, the latter is composed by 
the emission of numerous metal lines, mostly Fe\,II, that 
originates as a result of the absorption of 
ultraviolet photons generated in central parts of the envelope 
due to the reprocessing of the radioactive luminosity
(Xu \& McCray 1991). The scattering and absorption 
of the quasicontinuum in metal and hydrogen lines in outer layers 
produces broad P Cygni profiles, e.g., H$\beta$,
Fe\,II lines of the multiplet 42,  Na\,I doublet (Fig. 3).

A spectacular feature of the SN2004dj spectra, never observed 
before in SNIIP, is the strong blueshift of emission line maxima, 
most apparent in H$\alpha$ (Fig. 4). The profile of this line 
is clearly asymmetric with the maximum at $-1600$ km s$^{-1}$ 
in early (127 and 131 days) spectra. 
The blueshift decreases with time down to -600 km s$^{-1}$ on 
day 330. The blue peak, although weaker, 
is present also in H$\beta$. Remarkably, on day 131 the H$\alpha$
profile shows double horn structure with the red horn shift 
of $\approx 1600$ km s$^{-1}$. This indicates the 
bipolar, although asymmetric,  structure of the hydrogen excitation.

Among type IIP supernovae only two cases of significant
H$\alpha$ line asymmetry were mentioned earlier: SN1987A 
(Phillips \& Williams 
1991) and SN1999em (Elmhamdi et al. 2003a). In both cases the 
redshift was reported. The redshift in SN1987A
has been interpreted 
as a result of the ejection of significant amount of $^{56}$Ni in the 
far hemisphere in otherwise spherical envelope 
(Chugai 1991). The similar interpretation 
was proposed for the redshift of H$\alpha$ and HeI 10830 \AA\ 
lines in SN1999em (Elmhamdi et al. 2003a).
We adopt the same explanation for 
the H$\alpha$ asymmetry in SN2004dj with the only difference that 
in this case the $^{56}$Ni distribution is skewed towards the observer. 

The manifestation of the line asymmetry in SNIIP
is displayed in Fig. 5, which  shows H$\alpha$ and [Ca\,II] 
doublet in SN2004dj on day 149, SN1987A on day 198 
(Phillips \& Williams 1991) and SN1999em on day 140 
(Elmhamdi et al. 2003a). 
Apart from the sign of asymmetry, SN2004dj differ by 
the larger shift and the amplitude of the asymmetric H$\alpha$
component.
The [Ca\,II] doublet of SN2004dj differs also very much from 
that in other supernovae (Fig. 5). SN2004dj shows the striking blue
peak in [Ca\,II] at $-1600$ km s$^{-1}$ that is seen in H$\alpha$.
This peak reflects, presumably,
the local overexcitation of Ca\,II related to the asymmetry of 
$^{56}$Ni ejecta likewise in H$\alpha$. 

Other spectral differences between SN2004dj and SN1999em 
related to different signs of asymmetry are noteworthy. 
The absorption component of H$\alpha$ in SN2004dj is markedly 
weaker than in SN1999em (Fig. 6). This is caused by 
the presence of the blue emission peak in SN2004dj that  
fills in the absorption.  
Absorption component of H$\beta$, on the contrary, is stronger 
in SN2004dj. The reason for that is also the ejection of 
$^{56}$Ni primarily towards the observer. This causes 
higher hydrogen excitation in the near hemisphere of 
SN2004dj compared to SN1999em where most of $^{56}$Ni is shifted 
towards the far hemisphere. As a result the H$\beta$ optical depth 
in the near hemisphere of SN2004dj is greater than in SN1999em.

Interestingly, spectra of SN2004dj at the plateau epoch 
(Korcakova et al. 2005) 
demonstrate larger expansion velocities of the H$\alpha$ absorption 
minimum compared to SN1999em. On JD2453221 (day 51 for the assumed
outburst date) radial velocity of the 
H$\alpha$ absorption minimum is -6955 km s$^{-1}$
(Korcakova et al. 2005) compared to 
-5000 km s$^{-1}$ in SN1999em on 54 day (Elmhamdi et al. 2003a). 
We believe that higher velocity of SN2004dj reflects the 
higher degree of the hydrogen excitation in the near 
hemisphere rather then higher energy-to-mass ratio. 

Another remarkable peculiarity of SN2004dj is 
the quasicontinuum shape in the range of 
5000--5700 \AA\ which differ from that in SN1999em. Note, 
the quasicontinua of SN1999em and SN1987A are rather similar in 
the same range. The plausible explanation involves the different sign of
$^{56}$Ni asymmetry in SN2004dj and SN1999em. Actually, 
the $^{56}$Ni displacement should result in the different 
conditions for the quasicontinua formation in asymmetric 
hemispheres because of the different excitation degree. 

\subsection{H$\alpha$ luminosity and $^{56}$Ni mass}

The integrated H$\alpha$ flux is measured 
in the spectra which were calibrated using 
$V$ and  $Rc$ photometric data. 
The resulting H$\alpha$ flux is the average between fluxes derived 
from $V$ and $Rc$ calibrations. We omit spectra on days 131 and 237
which are
strongly affected by the atmospheric dispersion, and 
the spectrum on day 330 which cannot be calibrated confidently 
because the strong [Ca\,II] emission is missing. 
The obtained H$\alpha$ fluxes, uncorrected ($F$) and corrected 
($F^{\rm corr}$) 
for the dust absorption, are given in Table 3. The last column 
is the H$\alpha$ luminosity. The error of the flux (0.12 dex) 
is estimated as half of the maximal difference of fluxes derived 
from $V$ and $Rc$ calibrations. 

The H$\alpha$ luminosity at the nebular epoch is used here
to estimate $^{56}$Ni in SN2004dj likewise this was
done earlier for SNIIP (Elmhamdi et al. 2003b). 
The method suggests the computation of H$\alpha$ luminosity 
powered by the radioactive decay $^{56}$Co -- $^{56}$Fe. 
Three levels plus continuum 
hydrogen atom is adopted, although recombinations on the 
excited levels are included. The photoionization from 
the second and third level by continua is taken into account. Radiation 
transfer in continua is treated in the escape probability 
approximation (Chugai 1987). The two-zone model envelope consists of the 
macroscopically mixed core and outer 
envelope. The model parameters are the mass $M$, kinetic energy $E$, 
$^{56}$Ni mass, mass fraction in the core  
$f_{\rm mix}$ and the fraction of the core residing in 
metals and helium  $f_{\rm m}$. The quasicontinuum SED is 
described by the plankian spectrum with the color temperature 
$T_{\rm c}$. In contrast to the previous version here 
the electron temperature $T_{\rm e}$ is calculated from the 
energy balance which takes into account cooling in 
Ca\,II, Mg\,II, Fe\,II (approximately), O\,I lines,
$ff$-emission and adiabatic cooling.
We abandon the approximation of the homogeneous 
density assuming that the density in the core is three 
times larger than the outer shell density. This mimics
the density distribution in the real SNIIP envelope. 

The model takes into account the outcome of the 
macroscopic mixing of $^{56}$Ni in the exploding star with the 
stratified chemical composition. We adopt that fragments of 
the $^{56}$Ni material are embedded in cocoons composed of metals 
and helium, i.e., of the material lacking the hydrogen. The 
optical depth of the cocoon for gamma-rays is determined by the mass 
and the surface area of cocoons,
$\tau_{\rm c}= kf_{\rm m}f_{\rm mix}M/(4\pi f_{\rm s}R_1^2)$, 
where $k=0.03(1+X)$ cm$^{2}$ g$^{-1}$ (Fransson and Chevalier 1989),
$X$ is the hydrogen abundance, $R_1$ is the radius of the mixed core,
$f_{\rm s}$ is the mixing parameter. The latter is the ratio of the 
cocoon surface area to $4\pi R_1^2 $. Below we show results for 
$f_{\rm s}=2$ (moderate mixing) and $f_{\rm s}=20$ (strong mixing). 
The adopted ejecta mass is $M=13~M_{\odot}$, presumably the 
mass of SN1999em ejecta (Nadyozhin 2003) that corresponds to the 
$15~M_{\odot}$ main sequence mass. The kinetic energy is 
$E=10^{51}$ erg, the typical value for SNIIP (Nadyozhin 2003). 
Other parameters are $f_{\rm mix}=0.4$, $f_{\rm m}=0.3$, 
i.e., $\approx1.56~M_{\odot}$ of the core resides in metals and 
helium. The remainder of helium, from several tens to 
$\sim 1~M_{\odot}$ is presumably mixed with the hydrogen. 
Note, for the star with the initial mass $M\approx15~M_{\odot}$ 
the predicted mass of newly synthesised
 metals and helium in the supernova is about $3~M_{\odot}$ 
 (Woosley \& Weaver 1995), in agreement with our assumptions.

The optimal models of the H$\alpha$ luminosity in SN1999em at 
the nebular epoch $t\geq 130$ days are presented in Fig. 7 for 
the two values of the mixing parameter $f_{\rm s}$. The adopted 
$^{56}$Ni mass $M(^{56}{\rm Ni})=0.027~M_{\odot}$ is equal to 
the value recovered 
from the tail of $V$ light curve 
(Elmhamdi et al. 2003b). The color temperature of the 
quasicontinuum is the only free parameter. Its value 
turns out to be $T_{\rm c}=5500$~K for the model with $f_{\rm s}=2$ and 
$T_{\rm c}=5000$~K for $f_{\rm s}=20$.
 
For SN2004dj we adopt the same model as for SN1999em.
The only difference is the $^{56}$Ni mass, which is 
$M(^{56}{\rm Ni})=0.02~M_{\odot}$, 
the value recovered from the $V$ light curve. 
The plot (Fig. 7, lower panel) shows that the model of the 
H$\alpha$ luminosity with this amount of $^{56}$Ni fits 
to the observational data within errors. Of course, this 
fact cannot be considered 
as a proof that the $^{56}$Ni mass is actually $\approx0.02~M_{\odot}$
given uncertainties of other parameters.
On the other hand, this indicates that the $^{56}$Ni mass recovered 
from the $V$ light curve is approximately consistent with the 
H$\alpha$ luminosity.

\section{Line asymmetry and $^{56}$Ni distribution}

The line profile formed in the supernova envelope with the 
kinematic of the homologous expansion ($v=r/t$) depends 
on the distribution of the level populations and 
the continuum optical depth. At the nebular epoch 
the absorption in the quasicontinuum is weak in the 
red part of the spectrum ($\lambda>5000$ \AA). 
The profile of emission lines of H$\alpha$, 
[O\,I] 6300, 6364 \AA\ and [Ca\,II] 7291, 7324 \AA\ 
thus contain the information about the asymmetry of 
the radiation sources throughout the envelope including 
deep central zone. 

\subsection{[O\,I] 6300, 6364 \AA\ lines}

The lines in [O\,I] and [Ca\,II] doublets are 
partially  superimposed, so it would be reasonable to 
recover true line profile using representation 
of each line as a superposition of the minimal number 
of gaussian components. Each gaussian component is 
characterized by its radial velocity $v_{\rm r}$, 
Doppler width $w$ and relative amplitude $A$.

Oxygen doublet [O\,I] 6300, 6364 \AA\ is a convenient case for 
the decomposition because the velocity interval between lines
is large (3000 km s$^{-1}$).
Using only two gaussian components, central and blueshifted, 
we are able to reproduce successfully the doublet profile 
on day 237 and 330 (Fig. 8). Parameters 
of components (displacement, width, amplitude) are given in Table 4).
The line profiles of [O\,I] and H$\alpha$ on day 237 are 
quite similar (Fig. 4); on day 330, however, the difference is remarkable:
the [O\,I] maximum is at zero, while the H$\alpha$ maximum is 
blueshifted at $\approx-700$ km s$^{-1}$. This fact reflects, probably, 
the different distribution of hydrogen and oxygen. 

The increasing role of the central component can be explained 
in a picture of the two $^{56}$Ni components. 
Let the central component be embedded into the cocoon that 
is optically thick for gamma-rays and 
does not contribute in the [O\,I] doublet emission.
Note, the cocoon material may be even composed by oxygen 
and yet be unseen in [O\,I] lines. This is the case, if
formed CO and/or SiO molecules
cools the oxygen material (Liu \& Dalgarno 1995).
As the envelope expands, the cocoon gets optically thin and 
gamma-rays efficiently excite outer oxygen material thus 
increasing the role of the central component. A similar effect 
in H$\alpha$ will be studied below (Section 5.3).

For the doublet decomposition we used some value of the doublet ratio
$R=I(6300)/I(6364)$. The latter is equal to the ratio of line 
specific luminosities 
$h\nu_{ik}n_k A_{ki}\beta_{ik}$ [erg s$^{-1}$ cm$^{-3}$], 
where $\beta_{ik}$ is the escape probability, while the 
other notations are obvious. The optimal $R$ values
(Table 4, last column) are lower than nebular one, $R=3$, 
which implies that [O\,I] lines are optically thick. 
The similar situation takes place in other SNIIP 
(Spyromilio \& Pinto 1991; Chugai 1988). The doublet ratio is determined 
primarily by the oxygen concentration and only weakly depends 
on the electron temperature, since the excitation potential of the 
lower level of the transition of 6364 \AA\ is only 0.02 eV.
Since the oxygen is mostly neutral the ratio $R$ permits us 
to recover the oxygen density in the line forming zone. 
The observed ratio $R=1.33$ on day 237 leads to the 
oxygen number density 
$n(\mbox{O})=6.63\times10^9(237\,\mbox{d}/t)^3$ cm$^{-3}$, 
where $t$ in days and $T_{\rm e}=5000$ K is adopted. 
The density evolution $n({\rm O})\propto t^{-3}$
predicts the ratio $R=1.77$ on day 330, 
in good agreement with the value $R=1.8$ found from 
observations (Table 4). 
This confirms the adopted explosion date of SN2004dj, although  
the accuracy of the reconstruction of the 
explosion date from the ratio $R$ is not better than 
$\pm30$ days.

\subsection{[Ca\,II] 7391, 7324 \AA\ lines}

The lines of [Ca\,II] doublet also can be 
represented by two gaussian components, central and 
blueshifted (Fig. 9, Table 5).  However, the broad central component 
shows some deviation from gaussian shape and 
requires an additional weak narrow component (Fig. 9, inset). 
On day 237 when [Ca\,II] and [O\,I] doublets can be compared, 
the line profiles are roughly similar, although some differences 
are apparent. Particularly, the blue component of [Ca\,II] 
shows the same shift but 1.6 times more narrow. The central 
component, on the contrary, is 1.7 times broader than in [O\,I] case. 
The average radial velocity of [Ca\,II] line 
(e.g., 7291 \AA) is 
 -150 km s$^{-1}$ on day 237, i.e., lower than that of [O\,I]
 line at the same epoch (-770 km s$^{-1}$).  

These differences are likely caused by the formation of these 
lines in different zones. Actually, stellar evolution for 
massive stars (e.g., $15~M_{\odot}$) and supernova models  
predicts that Ca is synthesised in two processes: static 
oxygen burning in the presupernova and explosive oxygen 
burning during the shock wave propagation (Woosley \& Weaver 1995).
Calcium, therefore, is expected to reside in products of the 
oxygen burning (in Si/S matter), not in the oxygen matter. 
Moreover, the [Ca\,II] 7391, 7324 \AA\ lines may also form in 
hydrogen or helium material (Li \& McCray 1993) with the 
main sequence abundance.
 
Given uncertainty of the site of the [Ca\,II] doublet 
formation, it is interesting that the doublet ratio on day 149 is 
$R=I(7291)/I(7324)\approx 1.1$, i.e., lower than the 
nebular one ($R=1.5$). If this is caused by the optical depth 
effect, then the required number density of Ca should be
$n(\mbox{Ca})>3.6\times10^6(149\,{\rm d}/t)^3$ cm$^{-3}$.
Assuming that the [Ca\,II] doublet is emitted by the hydrogen matter 
we find the required density of hydrogen for normal Ca abundance 
$\rho > 3.7\times10^{-12}(149\,{\rm d}/t)^3$ g cm$^{-3}$. 
This should be compared with the oxygen density found 
from  oxygen doublet ratio,  
$\rho \approx 7\times10^{-13}(149\,{\rm d}/t)^3$ g cm$^{-3}$.
The latter turns out to be five time lower compared to the 
required hydrogen density, quite reverse to expectations. 
Actually, the analysis of SN1987A shows that the 
oxygen is distributed in dense condensations with the 
filling factor of $f\sim 0.1$ and the oxygen density 
is by order of magnitude greater than the 
average density (Spyromilio \& Pinto 1991; Chugai 1988). 
The found controversy indicates that the hypothesis of the 
[Ca\,II] doublet formation in the hydrogen material 
faces serious problems.
However, given uncertainty of the 
quasicontinuum behavior in the 7300 \AA\ band one cannot
preclude that we underestimated the doublet ratio. 
We thus do not rule out 
the formation of [Ca\,II] doublet in the hydrogen material 
although this possibility  seems doubtful.

\subsection{Modeling H$\alpha$ asymmetry}
 
Here we address
the H$\alpha$ line profile and its connection with 
the asymmetry of the $^{56}$Ni distribution.
We facilitate the physics of the energy cascade resulting in 
the H$\alpha$ emission and assume that the specific 
H$\alpha$ luminosity, $\eta=h\nu_{23}n_3A_{32}\beta_{23}$, 
is proportional to the local deposition rate 
$\eta=C\epsilon_{\rm d}$ with the 
constant factor $C$. To obtain the H$\alpha$ 
line profile one has to define the $^{56}$Ni distribution 
and to calculate the distribution of the deposition 
$\epsilon_{\rm d}$. 
The gamma-ray transfer is computed in a single flight 
approximation with the absorption coefficient adopted above 
(section 4.2).

The model envelope expands homologously ($v=r/t$) 
and has the spherical mass distribution (but $^{56}$Ni). 
The density is described by the exponential distribution 
$\rho=\rho_0\exp(-v/v_0)$ with $\rho_0\propto t^{-3}$.
The exponential law is fairly good for the 
bulk of the inner envelope where the power index 
$|d\ln \rho/d\ln v|$ gradually increases outwards 
(e.g., Utrobin 2004). The parameters $\rho_0$ and $v_0$ 
are determined by $M$ and $E$. We adopt as above 
$M=13~M_{\odot}$ and $E=10^{51}$ erg. 

The $^{56}$Ni distribution is set by 
three components: central homogeneous sphere and two 
cylindrical jets with the angle $\theta$ 
between the jet axis and the line of sight (Fig. 10). In the 
regions, where jets are superimposed on the central spherical component,
the density is assumed to be equal to the
jet density. We found that the opposite condition is less 
consistent with observations. 
The density of $^{56}$Ni in jets is assumed to decrease 
linearly outwards down to zero at the outer jet boundary.
The chemical composition of the envelope is homogeneous with the 
hydrogen abundance $X=0.7$. In the central zone of the radius 
$R_{\rm s}=v_{\rm s}t$ the matter distribution is presumably inhomogeneous 
with the filling factor of the hydrogen matter linearly 
decreasing inwards from unity at $R_{\rm s}$ to zero 
in the center. This assumption is consistent 
with intuitive picture of the mixing of helium and metals
 with the hydrogen envelope.

The study of the parameter space of the model led us 
to conclude that only the version with 
cocoons around $^{56}$Ni material is able to reproduce 
the evolution of H$\alpha$ profile.
The cocoon model has been already used above to describe 
the evolution of H$\alpha$ luminosity. Here this 
model should be refined given more complicated geometry of mixing.
We assume that $^{56}$Ni is represented by an ensemble 
of spherical fragments each coated by cocoon of the 
optical depth $\tau_{\rm c}$ for gamma-rays. 
The average gamma-ray emissivity is then 
multiplied by the attenuation factor $\exp(-\tau_{\rm c})$.
We adopt $\tau_{\rm c}=const$ 
in the central zone of the radius $R_{\rm s}$, while in jets 
outside the central zone, the cocoon optical depth decreases 
with the velocity $v$ as 
\begin{equation}
\tau_{\rm c}=\tau_0\left(\frac{v_k-v}
{v_k-v_{\rm s}}\right)^p\,,
\end{equation}
where $v_k$ is the outer velocity $k$-th jet ($k=1, 2$).
The adopted power index is $p=3$. 
The parameter $\tau_0$ may be written via the cocoon mass of 
$M_{\rm c}$ as 
$\tau_0=kM_{\rm c}/4\pi f_{\rm s}R_{\rm s}^2$ where 
$f_{\rm s}$ is the mixing parameter introduced earlier 
(Section 4.2). The decrease of $\tau_{\rm c}$ outwards 
corresponds to the increase of the mixing degree. This 
is just what is expected from the mixing physics: 
the material propagated through the larger mass should 
experience larger mixing degree.

Parameters of the optimal model, i.e., the angle $\theta$ between 
the line of sight and the axis of $^{56}$Ni, the expansion 
velocity  of the central component $v_{\rm s}$, boundary velocity of 
jets ($v_1$, and $v_2$), the mass of jets, and cocoon mass for 
the mixing parameter $f_{\rm s}=2$ are given in Table 5.
The accuracy of the determination of velocities is 
about 20\%, angle $\theta$ is determined with the accuracy of 
$10^{\circ}$.  Interestingly, while the model of the 
H$\alpha$ luminosity does not impose strict constraints on $f_{\rm s}$,
 we find that assuming a
cocoon mass $M_{\rm c}\sim 1~M_{\odot}$  requires a moderate 
mixing, $f_{\rm s}\sim2$.

Model profiles of H$\alpha$ with and without cocoon are 
plotted along with observations in Fig. 11. The model with 
cocoon is apparently more adequately describes profile 
and its evolution during 131--330 days. 
Note, within the same model we are able to describe 
the non-trivial change from asymmetric two-horn 
profile on day 131, through the plateau plus blue peak 
on day 149, towards smooth asymmetric profile with the blueshifted 
maximum. 

The specific features of the recovered $^{56}$Ni distribution are 
the bipolar structure and the deviation from the point symmetry.
The mass of the front $^{56}$Ni jet is 
 comparable to the central component and 
 twice as larger compared to the rear $^{56}$Ni jet (Table 5).
Note, there is no need to admit the asymmetry of hydrogen 
distribution, which otherwise would require extremely large
degree of the supernova asymmetry. 

The one-dimension projection of the recovered  $^{56}$Ni distribution 
on the line of sight and the jet axis are shown in Fig. 12.
For comparison we show also 
the recovered line profile of [Ca\,II] 7291 \AA\ on day 149 
(Fig. 9), when the gamma-ray transfer effect is small and, therefore, 
the sources in [Ca\,II] lines follow the distribution 
of $^{56}$Ni. While there is a general correspondence 
between [Ca\,II] line  profile and $^{56}$Ni distribution we 
see apparent differences. The $^{56}$Ni projection on the line of sight 
does not show strong central peak that is present in [Ca\,II] line. 
This disparity indicates that the Ca distribution deviates from 
$^{56}$Ni distribution.

Interestingly, we found that the major mass of $^{56}$Ni should be 
isolated from the hydrogen by the cocoon that does not contain 
any hydrogen. Moreover, the mixing degree that determine the 
the thickness of cocoon should not be large. The similar 
property of the $^{56}$Ni distribution has been revealed 
by Liu \& Dalgarno (1995) in SN 1987A. To account for the 
evolution of CO 2.3 mcm emission they introduced 
a variable reduction factor for gamma-rays that decreases with time. 
This factor, we believe, reflects the existence of 
cocoons around $^{56}$Ni fragments in SN1987A as well.

\section{Discussion and conclusion}

We have presented results of the photometry and spectroscopy of 
SN2004dj, the usual type IIP supernova.
We have estimated $^{56}$Ni mass from the comparison of light 
curves of SN2004dj and SN1987A and from the modeling of 
H$\alpha$ luminosity. Both estimates suggests the 
$^{56}$Ni mass of $\approx0.02~M_{\odot}$. This is 3.8 times 
less than in SN1987A and comparable with $^{56}$Ni mass in 
SN1999em ($0.02-0.027~M_{\odot}$,  Elmhamdi et al. 2003a, 2003b).

The principal result of spectroscopic study of SN2004dj is 
the detection of a strong blueshift of the H$\alpha$ line 
 that is interpreted as a result of 
the asymmetric ejection of $^{56}$Ni in the 
spherically-symmetric envelope. 
The modeling of the H$\alpha$ line profile led us to conclude 
that the structure of the profile and its evolution are
well reproduced in the model of the asymmetric bipolar 
$^{56}$Ni ejecta with the two-fold difference of jet masses and 
the comparable masses of the massive jet and the central component.

We have found signatures of the bipolar structure in H$\alpha$ 
and [Ca\,II] line of SN1999em, which indicate that 
the asymmetric bipolar $^{56}$Ni ejecta is not exception for 
SNIIP.  Although in SN1987A 
spectral signatures of the bipolar structure are absent, 
Wang et al. (2002) see bipolar structure of inner 
region of SN1987A in images taken by {\em HST}.
The detection of bipolar structure of $^{56}$Ni ejecta 
at least in two normal SNIIP implies that bipolar explosion 
of the core-collapse supernovae can occure
for a moderately rotating red supergiant core. 

The significant deviation from the point symmetry in the $^{56}$Ni 
distribution revealed by at least three supernovae, 
namely SN1987A, SN1999em and 
SN2004dj, suggests that the total momentum may be 
essentially non-zero for synthesised iron-peak ejecta in SNIIP. 
In this case the finite momentum is presumably compensated 
by the envelope or/and neutron star.
According to recent study of proper motions of pulsars 
(Hobbs et al. 2005) neutron stars at birth acquire 
high velocities about 400 km s$^{-1}$ in average. These 
velocities unlikely could be explained by the sling effect of 
supernova in close binary, so the kick is the preferred explanation. 
The asymmetry of $^{56}$Ni ejecta in SNIIP combined 
with high pulsar velocities is thus a strong argument that the 
explosion of core-collapse supernovae is accompanied by the 
violation of the point symmetry. 

At present the likely mechanism for the asymmetry in core-collapse 
supernovae is the large-scale neutrino convection 
(Herant et al. 1992; Scheck et al. 2004), although the 
ejection of the supernova envelope with the 
required kinetic energy of $\sim10^{51}$ erg remains problematic 
(Buras et al. 2003). 

The detection of the bipolar structure of the $^{56}$Ni ejecta 
in normal SNIIP raises another intriguing question, whether the bipolar 
structure could emerge in the neutrino-driven explosion mechanism?
This problem was recently  
discussed in connection with the interpretation of the 
bipolar ejecta of Cas A  and SN1987A (Burrows et al. 2005). 
Authors considered the role of the moderate 
rotation in the creation of jets within 
neutrino-driven mechanism.  
However, it should be emphasised, that both these supernovae might in 
principle originate from presupernovae with 
the relatively fast rotation: Cas A is SNIb/c, and the 
nebular ring around SN1987A indicates fast rotation too.
As to SN2004dj and SN1999em, which are  normal SNIIP,  
here we deal with cores of red supergiants 
that are presumably slow rotators.  Therefore, 
it might be well that the origin of the 
asymmetric bipolar $^{56}$Ni ejection in SNIIP has 
nothing to do with the rotation.  

\bigskip 

We are grateful to V.L.~Afanasiev, N.B.~Borisov nad S.S. Kaisin 
for the help in observations. 
The work is partially supported by RFBR grants 4-02-17255, 
03-02-16341 and 04--2-16349.

{}

\clearpage

\begin{table}
  \caption{SN2004dj photometric data }
  \bigskip
  \begin{tabular}{lcccc}
  \hline

JD2450000+  &     $B$  &   $V$  &  $Rc$    & CCD \\

\hline

1929.4831 & 18.337 & 17.836   & 17.434 & K  \\
3321.6191 & 15.949 &  14.727  & 14.110 & E   \\
3355.5434 & 16.100 & 15.068   & 14.210 & VA \\
3357.5453 & 16.140 & 15.082   & 14.210 & VA  \\
3358.5417 & 16.116 & 15.083   & 14.187 & VA \\
3361.5745 & 16.128 & 15.129   & 14.243 & VA \\
3385.3628 & 16.320 & 15.342   & 14.311 & E  \\
3386.3845 & 16.346 & 15.345   & 14.326 & E  \\
3387.4160 & 16.320 & 15.351   & 14.320 & E  \\
3436.4913 & 16.534 & 15.666   & 14.662 & E  \\
3437.2683 & 16.539 & 15.689   & 14.678 & E  \\
3500.4002 & 16.044 &          &        & E  \\

\hline
\end{tabular}
\end{table}

\begin{table}
  \caption{Log of spectral observations}
  \bigskip
  \begin{tabular}{lcccc}
  \hline

Date  &   JD2450000+  &     Spectrograph  &  Range & Resolution \\
        &              &                   &  \AA\     &  \AA\ \\

 \hline      

{\rm 18.10.04} & 3298.5  &    LAT Scorpio & 3700 - 7500 & 13\\
{\rm 22.10.04} & 3301.5   &    1-m  UAGS  &   4000 - 7800  & 7.6\\
{\rm 09.11.04} & 3320.3   &   LAT Scorpio  &  3900 - 7500  & 13\\
{\rm 15.11.04} & 3325.3   &    LAT MPFS   &  4000 - 7000  & 6\\
{\rm 16.01.05} & 3387.2   &    LAT MPFS    &  4000 - 7000  & 6\\
{\rm 05.02.05} & 3406.8   &  LAT Scorpio    & 3900 - 7500 & 13\\
{\rm 09.05.05} &  3500.3   & LAT Scorpio    & 5700 - 7400  & 6\\
{\rm 07.06.05} &  3529.4  &  LAT Scorpio & 4000 - 5700  & 6\\

\hline
\end{tabular}
\label{t-par}
\end{table}

\clearpage
\begin{table}
  \caption{Flux and luminosity of H$\alpha$}
  \bigskip
  \begin{tabular}{lcccc}
  \hline

Date &  $t$  &   $F$    &   $F^{\rm corr}$       &        $L$   \\
     &  days  & \multicolumn{2}{c}{erg s$^{-1}$ cm$^{-2}$} 
      & erg s$^{-1}$\\

\hline

18.10.04  & 127 & $2.37\times10^{-12}$ & $2.72\times10^{-12}$ & 
$ 3.19\times10^{39}$\\
09.11.04  & 149 & $1.82\times10^{-12}$ & $2.09\times10^{-12}$ &
$ 2.45\times10^{39}$\\
15.11.04  & 155 & $1.84\times10^{-12}$ & $2.12\times10^{-12}$ &
$2.49\times10^{39}$\\
16.01.05  & 217 & $2.27\times10^{-12}$ & $2.61\times10^{-12}$ &
$3.07\times10^{39}$\\

\hline
\end{tabular}
\label{t-par}
\end{table}

\begin{table}
  \caption{Parameters of components of lines of [O\,I] and [Ca\,II] doublets}
  \bigskip
  \begin{tabular}{lcccccc}
  \hline

Doublet &  Age  & Component & $v_{\rm r}$  &  $w$  &  $A$  & $R$ \\
       &  days      &          & \multicolumn{2}{c}{km s$^{-1}$} &   &   \\

\hline

 [O\,I]   & 237  &   1      &  -1380   & 730   & 1.55   & 1.33\\

 [O\,I]   & 237  &   2      &  -30     & 750   & 1.21   & 1.33\\

 [O\,I]   & 330  &   1      &  -1200   & 750   & 1.2    & 1.8 \\ 

 [O\,I]   & 330  &   2      &    60    & 750   & 1.9    & 1.8 \\ 

 [Ca\,II] & 149  &   1      &  -1500   &  440  & 9.5    & 1.1 \\ 

 [Ca\,II] & 149  &   2      &    200   & 1300  & 0.53   & 1.1 \\ 

 [Ca\,II] & 237  &   1      &  -1270   & 470   & 8.8    & 1.3 \\ 

 [Ca\,II] & 237  &   2      &   200     & 1300  & 1.3    & 1.3 \\

\hline
\end{tabular}
\label{t-par}
\end{table}

\begin{table}
  \caption{Parameters of $^{56}$Ni distribution}
  \bigskip
  \begin{tabular}{lcccccc}
  \hline

 $\theta$ & $v_{\rm s}$  &  $v_1$ & $v_2$ &  $M_1/M_{\rm s}$ & 
 $M_2/M_{\rm s}$ & $M_{\rm c}$  \\
  &\multicolumn{3}{c}{km s$^{-1}$} &  &  & $M_{\odot}$  \\

\hline

 $30^{\circ}$ & 1400     & 2700  & 3500  &  1.07   &  0.49  & 1.3  \\

\hline
\end{tabular}
\label{t-par}
\end{table}

\clearpage

 \begin{figure}
\plotone{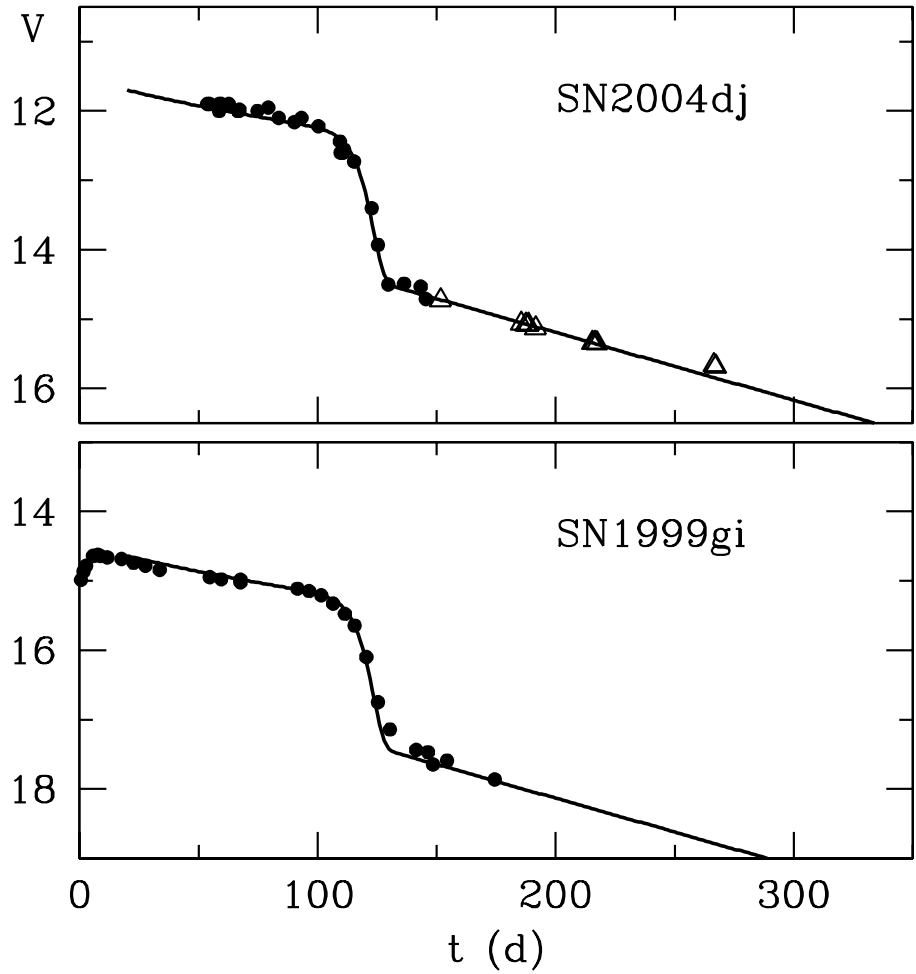}
\caption{ Light curves in $V$ band of SN2004dj and SN1999gi.
 {\em Triangles} are our data, {\em points} are amateurs data. 
 Solid line shows the analytical fit, the same in 
 both cases. The time is counted from the adopted explosion moment 
 (2004 June 13).}
\end{figure}
\clearpage
 \begin{figure}
\plotone{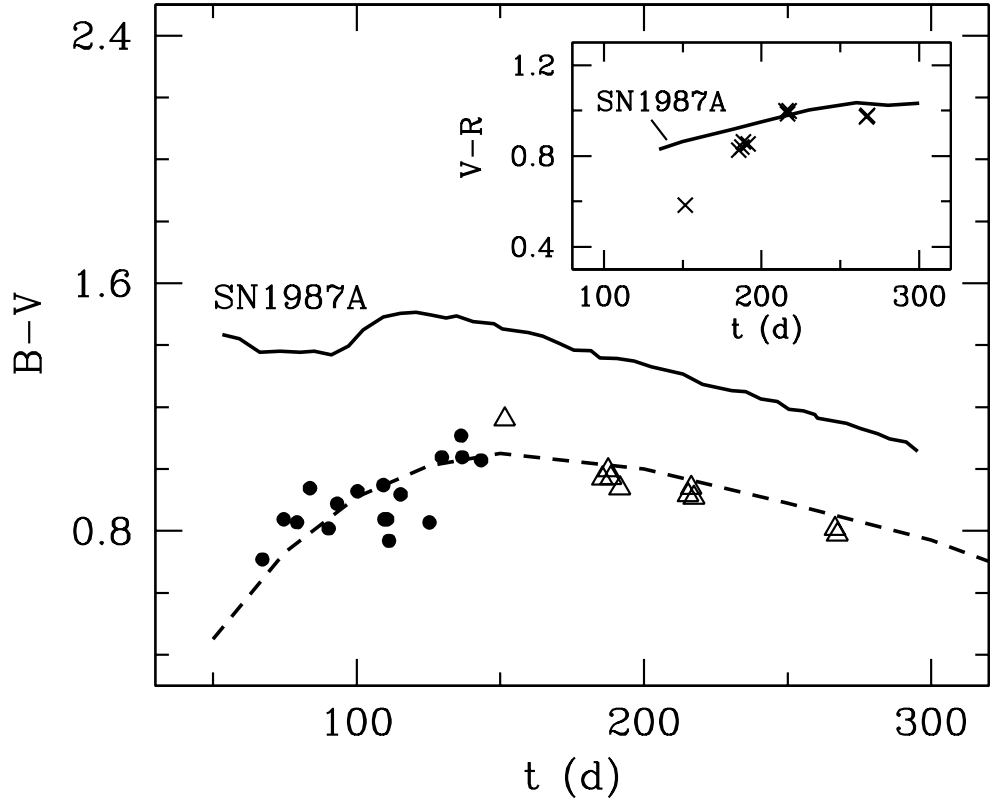}
\caption{Color $B-V$ of SN2004dj. 
{\em Triangles } are our data, {\em points} are amateurs data.
{\em Solid } line is the SN1987A color, 
{\em dashed} line is the lower envelope of $B-V$ colors of 
 SNIIP (Elmhamdi et al. 2003b). Inset shows the 
 $V-R$ color of SN2004dj ({\em crosses}) in comparison with 
 the color of SN1987A ({\em line}).
}
\end{figure}
\clearpage
 \begin{figure}
\plotone{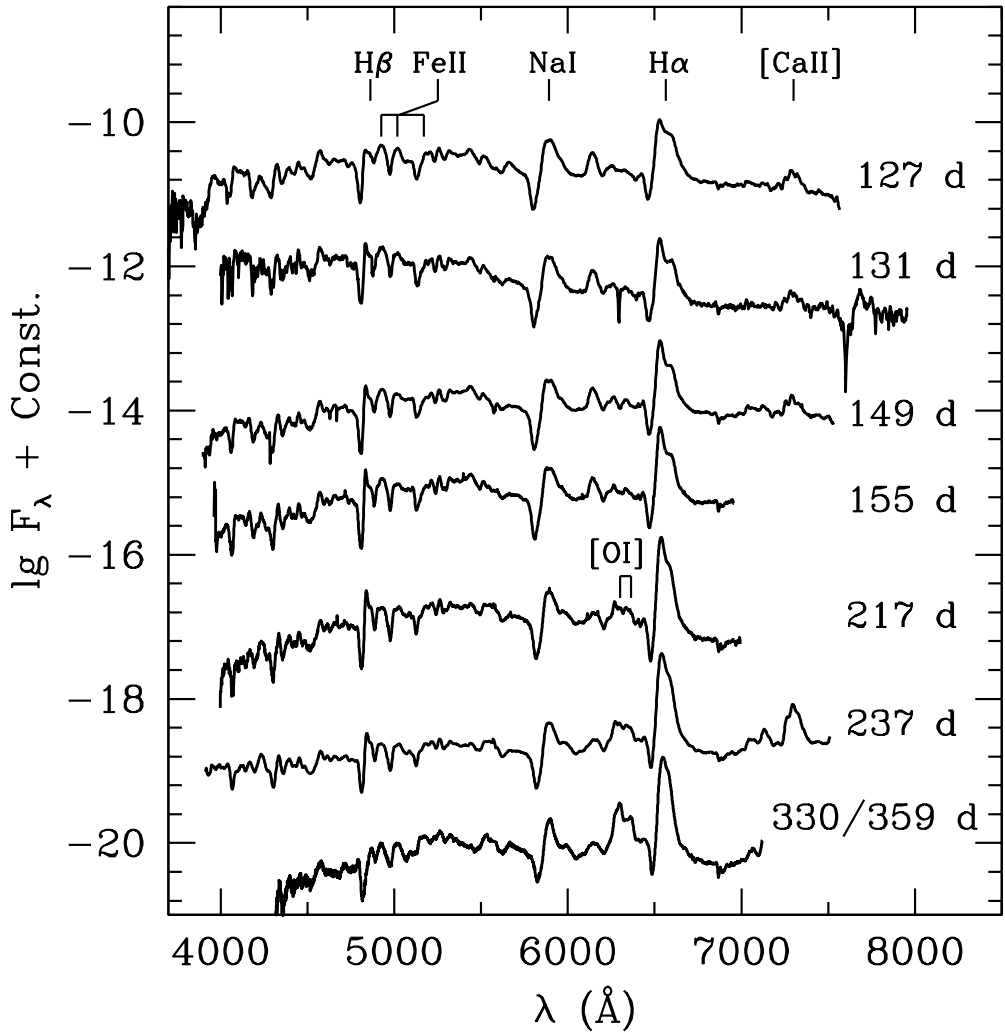}
\caption{Spectra of SN2004dj at different age and major 
identifications. The wavelenth is corrected for the galaxy redshift.
The age next to spectrum is measured from the explosion (2004 June 13).
}
\end{figure}
\clearpage
 \begin{figure}
\plotone{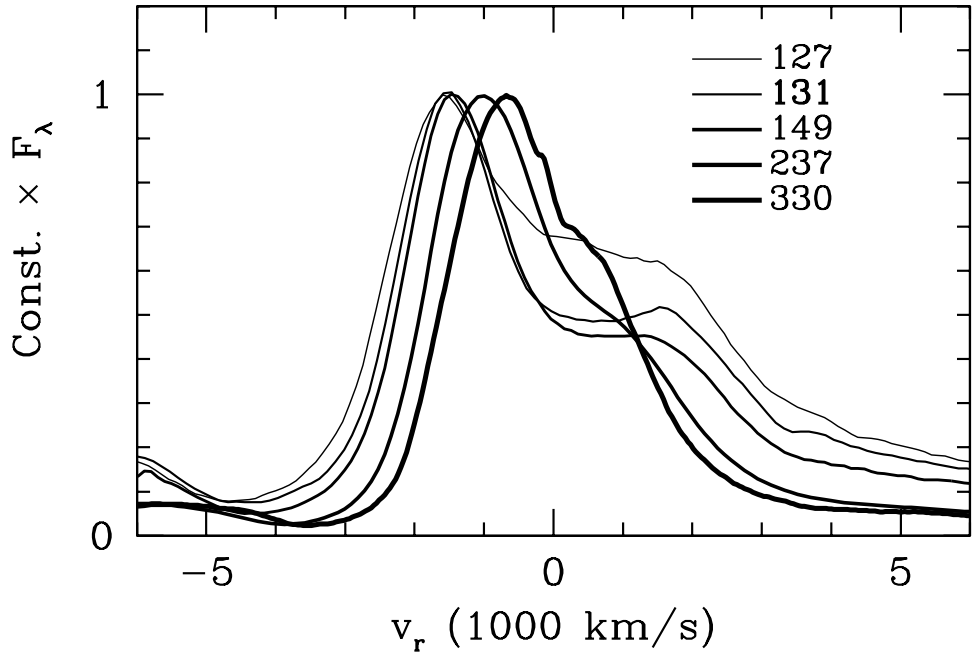}
\caption{Evolution of H$\alpha$ in spectra of SN2004dj.
The age is indicated next to line of a corresponding width.
}
\end{figure}
\clearpage
 \begin{figure}
\plotone{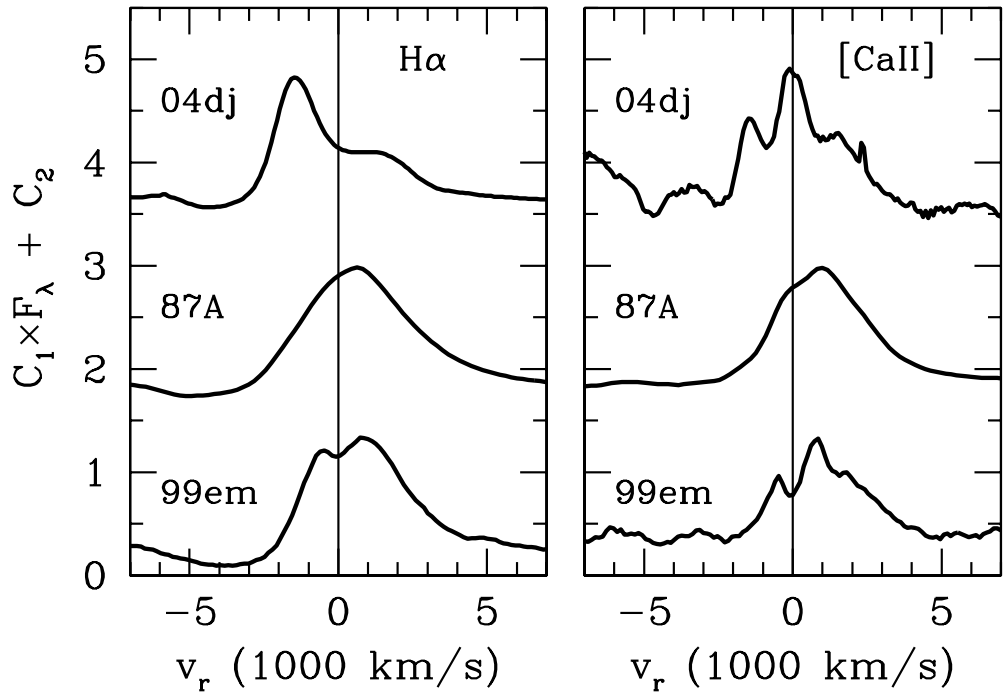}
\caption{H$\alpha$ line and [Ca\,II] doublet in 
spectrum of SN2004dj on day 149, in spectrum of SN1987A on day 198 
and in spectrum of SN1999em on day 140. The radial velocity 
of [Ca\,II] is given for the line of 7291 \AA.
}
\end{figure}
\clearpage
 \begin{figure}
\plotone{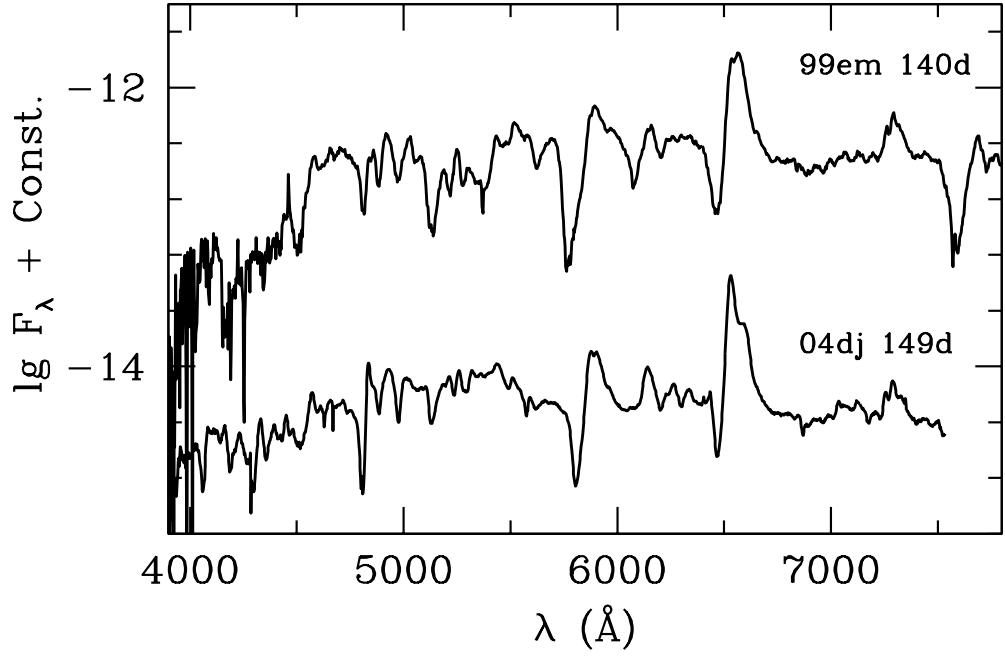}
\caption{Spectra of SN2004dj and SN1999em at similar phases.
Note, the difference of profiles and quasicontinua around 5600 \AA.
}
\end{figure}
\clearpage
 \begin{figure}
\plotone{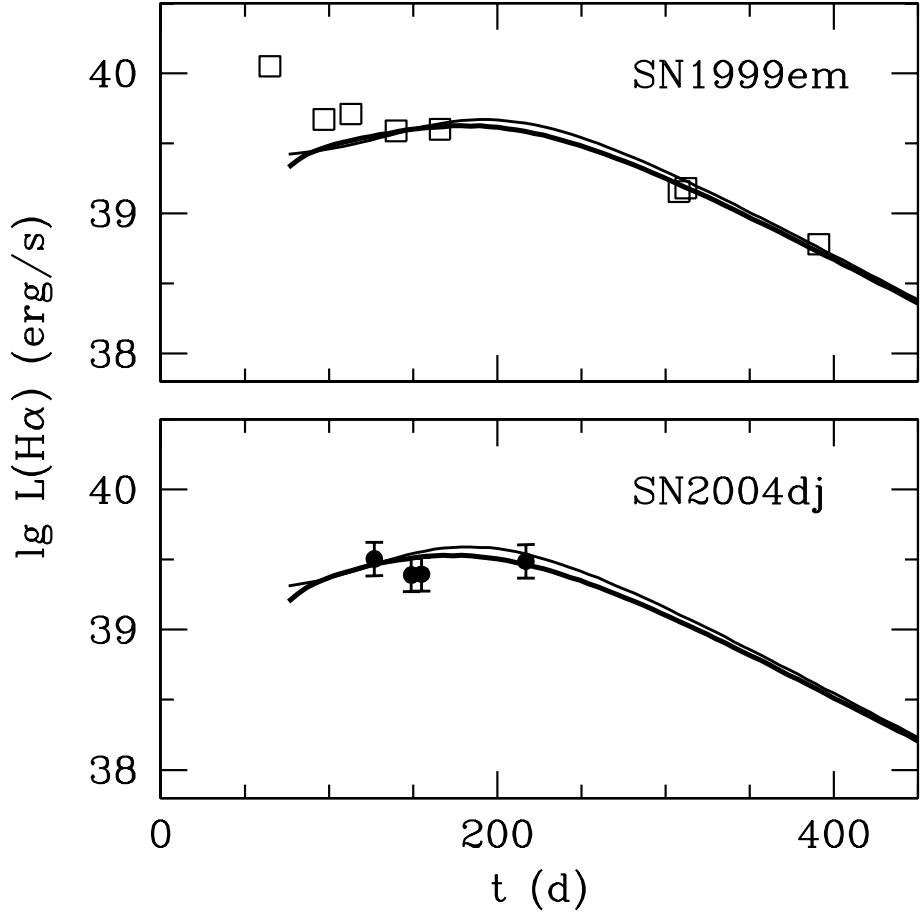}
\caption{Evolution models of H$\alpha$ for 
SN1999em and SN2004dj. Data for SN1999em (upper panel, {\em squares})
are taken from Elmhamdi (2003b). Model of SN1999em suggests
$^{56}$Ni mass of $0.027~M_{\odot}$ and two options of 
mixing parameter $f_{\rm s}=2$ ({\em thick} line) and
$f_{\rm s}=20$ ({\em thin} line). For 
SN2004dj the model is the same as for SN1999em, but 
$^{56}$Ni mass is $0.02~M_{\odot}$.
}
\end{figure}
\clearpage
 \begin{figure}
\plotone{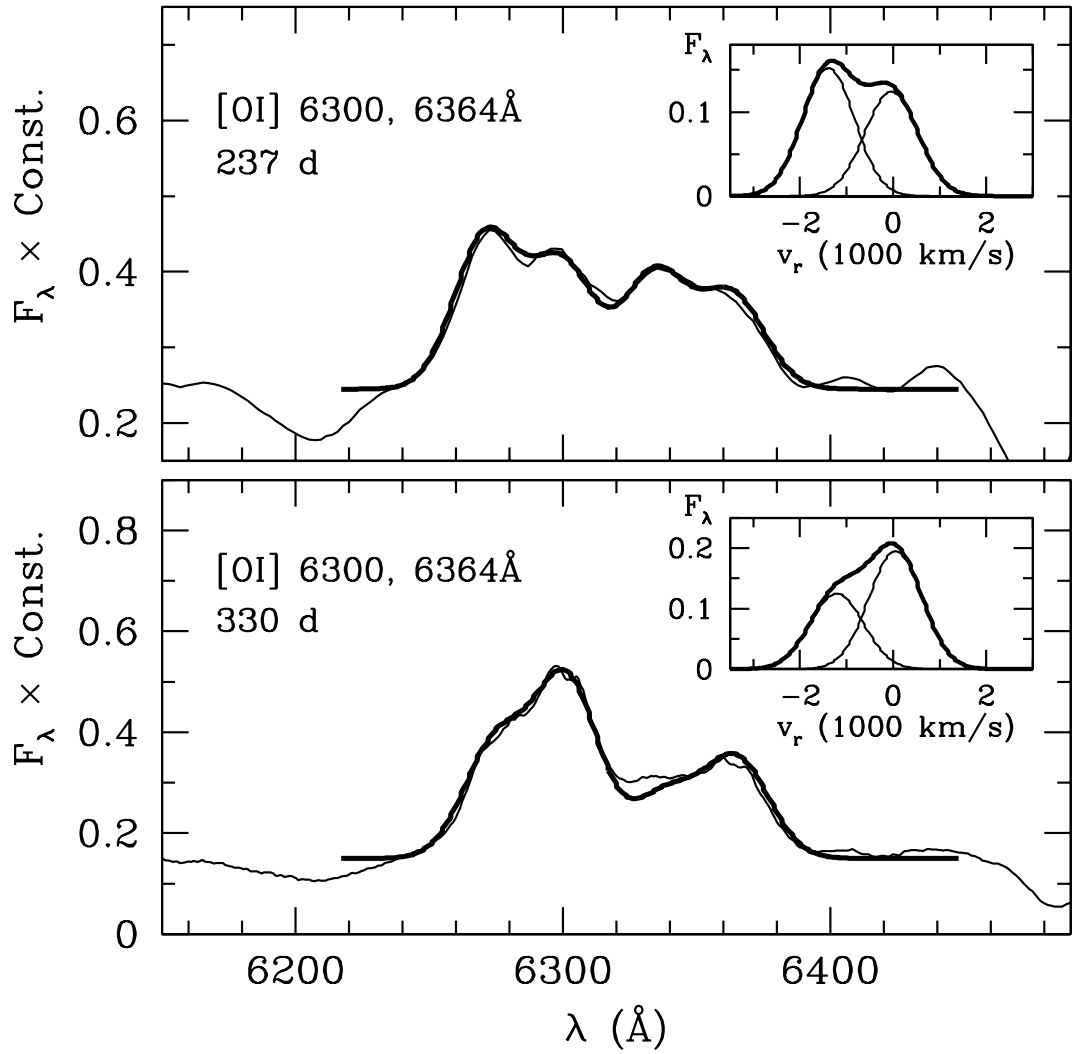}
\caption{Reconstruction of the line profile of 
[O\,I] 6300, 6364 \AA\ doublet in SN2004dj on day 237 and 330. 
{\em Thick} line is the model,  {\em thin} line is observation. 
Inset shows model line profile ({\em thick} line) along with 
gaussian components. 
}
\end{figure}
\clearpage
 \begin{figure}
\plotone{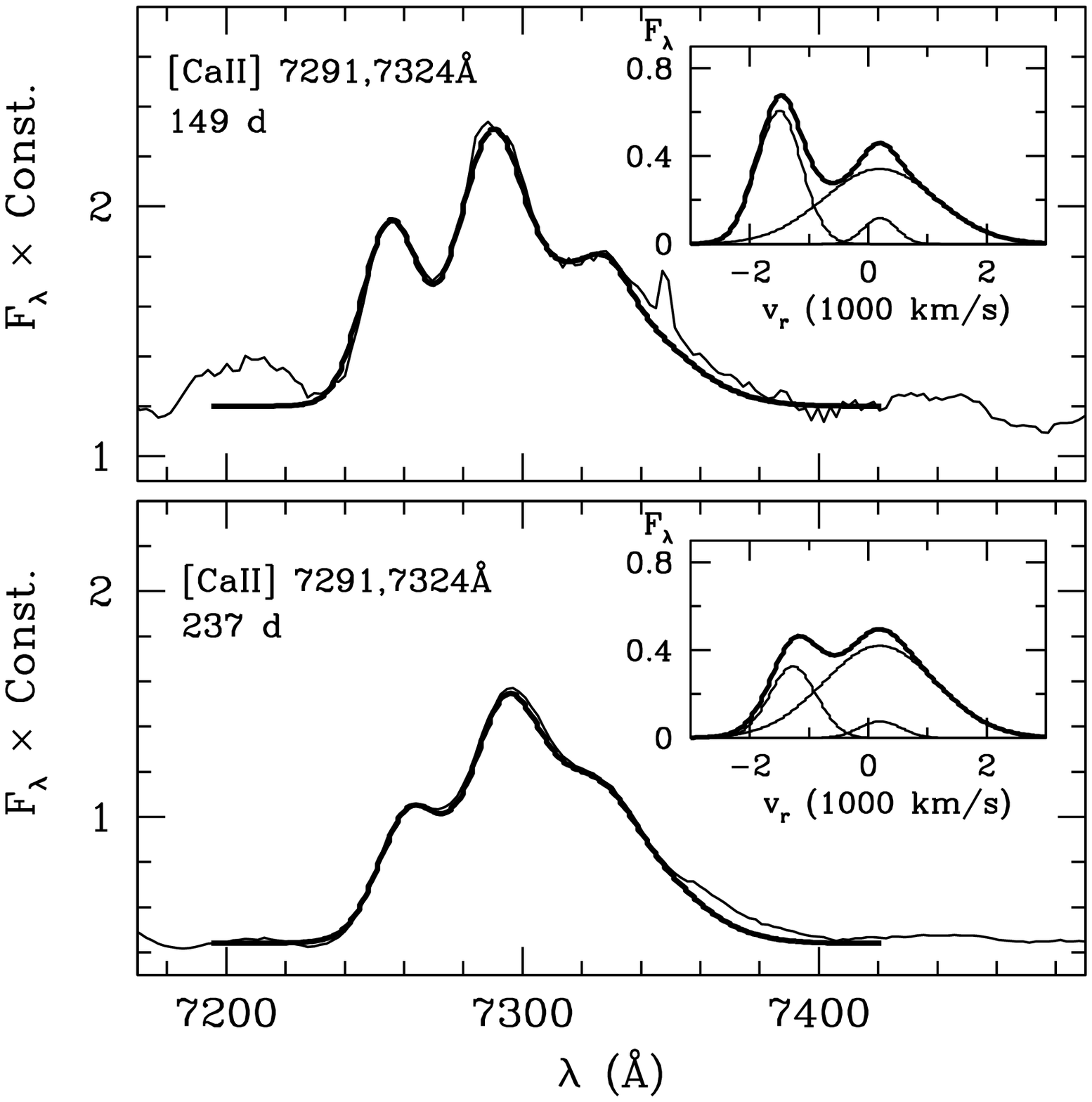}
\caption{Reconstruction of the line profile of 
[Ca\,II] 7291, 7324 \AA\  doublet in SN2004dj on day 149 and 237.
{\em Thick} line is the model,  {\em thin} line is observation. 
Inset shows model line profile ({\em thick} line) along with 
gaussian components. 
}
\end{figure}
\clearpage
 \begin{figure}
\plotone{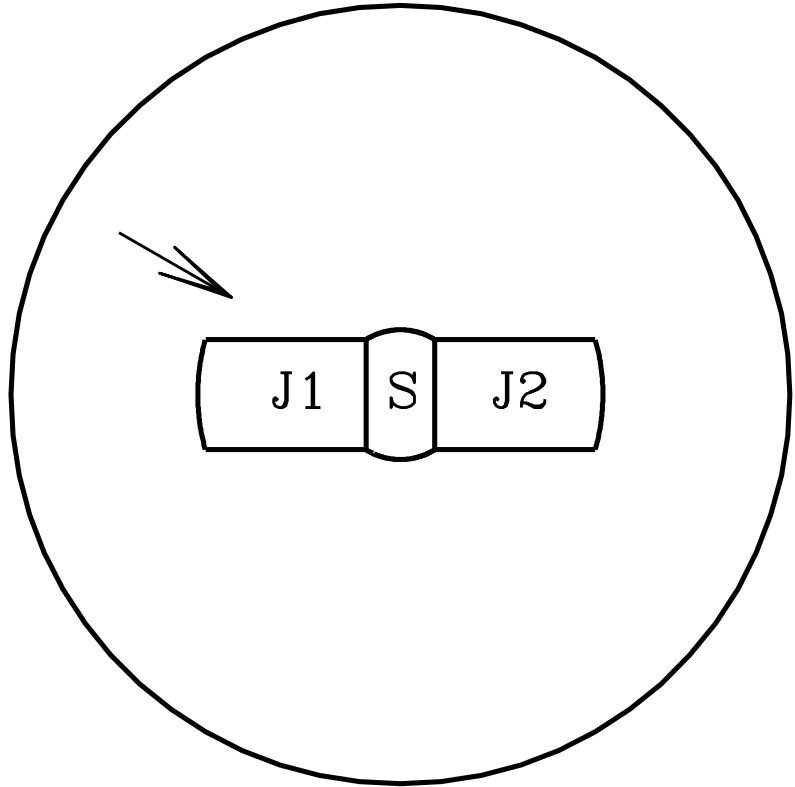}
\caption{Schematic representation of the spherical 
envelope of SN2004dj with bipolar $^{56}$Ni distribution,
The $^{56}$Ni distribution includes the central component 
S and two jets J1 and J2. Arrow shows 
the line of sight. 
}
\end{figure}
\clearpage
 \begin{figure}
\plotone{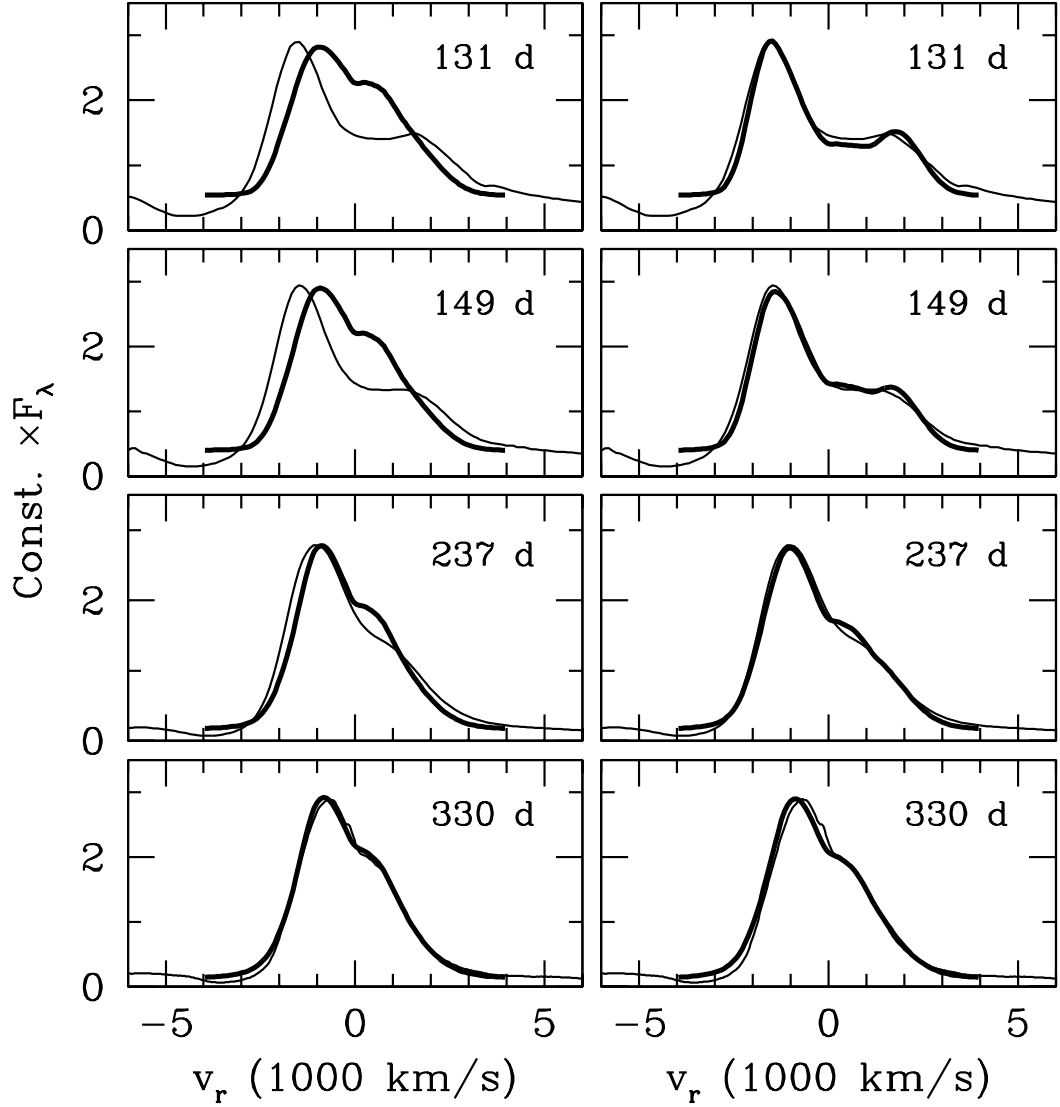}
\caption{ Models of the H$\alpha$ 
in SN2004dj with the asymmetric bipolar $^{56}$Ni 
distribution. {\em Thick} line is the model, {\em thin} line is 
observations. Models of $^{56}$Ni distribution 
without cocoon ({\em left} panels) and with cocoon ({\em right} 
panels) are shown. 
}
\end{figure}
\clearpage
 \begin{figure}
\plotone{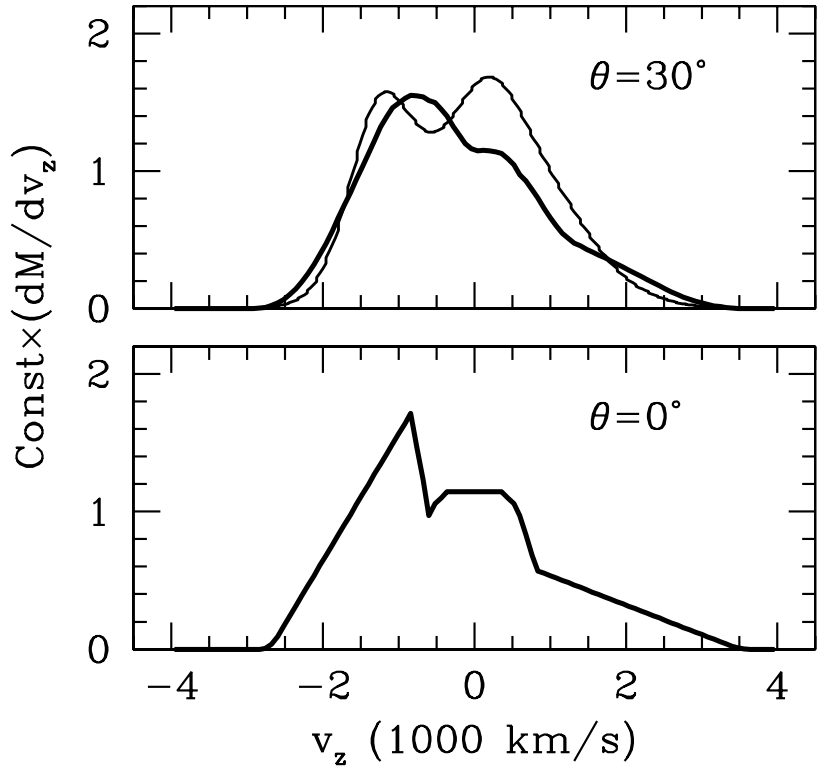}
\caption{Mass distribution of $^{56}$Ni along the radial velocity.
Upper panel shows projection on the line of sight 
({\em thick} line) together with the model profile of 
[Ca\,II] 7291 \AA\ line on day 149 (Fig. 9). The lower 
panel shows the projection of $^{56}$Ni distribution on the 
polar axis of ejecta.
}
\end{figure}

\end{document}